\documentclass[twocolumn,showpacs,prl]{revtex4}
\usepackage{times,xspace}
\usepackage{amsbsy,amssymb,amsmath,bm}
\usepackage{graphicx,color,epsfig,rotate}
\usepackage{fancyhdr}

\def\bbbc{{\mathchoice {\setbox0=\hbox{$\displaystyle\rm C$}\hbox{\hbox
to0pt{\kern0.4\wd0\vrule height0.9\ht0\hss}\box0}}
{\setbox0=\hbox{$\textstyle\rm C$}\hbox{\hbox
to0pt{\kern0.4\wd0\vrule height0.9\ht0\hss}\box0}}
{\setbox0=\hbox{$\scriptstyle\rm C$}\hbox{\hbox
to0pt{\kern0.4\wd0\vrule height0.9\ht0\hss}\box0}}
{\setbox0=\hbox{$\scriptscriptstyle\rm C$}\hbox{\hbox
to0pt{\kern0.4\wd0\vrule height0.9\ht0\hss}\box0}}}}

\newcommand{\ignore}[1]{}
\newcommand{\mComment}[1]{}
\newcommand{\gComment}[1]{}
\newcommand{\jComment}[1]{}
\newcommand{\rComment}[1]{}
\newcommand{\lComment}[1]{}

\renewcommand{\mComment}[1]{\textcolor{blue}{Bruce: #1}}
\renewcommand{\gComment}[1]{\textcolor{red}{Zohar: #1}}
\renewcommand{\jComment}[1]{\textcolor{green}{Cristian: #1}}

\begin{document}

\preprint{APS/123-QED}

\title{Stability of spontaneous quantum Hall state in the Triangular Kondo-lattice model}

\author{Yasuyuki Kato}

\author{Ivar Martin}

\author{C. D. Batista}

\affiliation{
	Theoretical Division, Los Alamos National Laboratory, Los Alamos, New Mexico 87545, USA
}%

\date{\today}

\begin{abstract}
We study the  behavior of the quarter-filled Kondo lattice model on a triangular lattice by combining a zero-temperature variational approach and finite-temperature  Monte-Carlo simulations.
For intermediate coupling between itinerant electrons and classical moments ${\bf S}_j$, we find a thermodynamic phase transition into an
exotic spin ordering with uniform scalar spin chirality and $\langle {\bf S}_j \rangle=0$. The state exhibits spontaneous quantum Hall effect. We also study how its properties are affected by application of an external magnetic field.

\end{abstract}

\maketitle

The very broad spectrum of physical phases and responses of strongly correlated materials is 
rooted in the simple fact that electrons carry both {\it charge} and {\it spin}. The interplay between these elementary
degrees of freedom gives rise to unconventional forms of superconductivity, multiferroic behavior, 
giant magnetoresistance, and heavy fermion physics. It also entails the possibility of strong magneto-electric effects.
In insulators, the electric polarization can be induced or modified by applying a magnetic field, or by coupling to certain magnetic orderings \cite{tokura2006}.  
In metallic systems, the magneto-electric effects are manifested in the conductivity tensor. The giant \cite{Baibich1988,Binasch1989} and colossal \cite{lalena2010book} magneto-resistance effects are examples in which the diagonal part of the conductivity tensor is dramatically modified. In addition, magnetic ordering can lead 
to changes in the off-diagonal components of the conductivity tensor. For instance, the   ``anomalous Hall effect'' observed in the presence of ferromagnetic order, even in the absence of externally applied magnetic field, ${\bf H} = 0$, has been a subject of active research over the past several decades \cite{sinitsyn2008,nagaosa2010,xiao2010}.

Recently, a residual Hall effect was observed in the metallic pyrochlore system Pr$_2$Ir$_2$O$_7$, in zero magnetic field and in the absence of uniform magnetization \cite{Machida2010}. 
This observation strongly suggests that the effect is caused by a magnetic structure with non-zero average scalar spin chirality, which for a single triangular plaquette is  $\langle \chi_{ijk} \rangle = \langle {\bf S}_i \cdot {\bf S}_j \times {\bf S}_k \rangle$ \cite{ohgushi2000}. 
Scalar spin chirality breaks time reversal and parity symmetries and can be stabilized even in the absence of 
usual magnetic ordering: $\langle {\bf S}_i \rangle=0$ \cite{domenge2005}. The symmetry properties of this order parameter
lead to unusual magneto-electric effects both in metals \cite{ohgushi2000, martin2008,shindou2001,taguchi2001,neubauer2009,lee2009} and in insulators \cite{lev2008}. 
The origin of the magneto-electric coupling lies in the Berry phase \cite{berry1984} that electrons accumulate as they traverse closed paths in the real space. 
The Berry phase is equal to half of the solid angle subtended by the electron spin as it moves around the path; e.g.,  in the continuum limit of smooth magnetic textures, it is proportional to the integral of the scalar spin chirality (or Berry curvature) over the area enclosed by the loop.
For a given spin species (locally parallel or antiparallel to the magnetic texture), the effect of the Berry phase is equivalent to an orbital magnetic field, and therefore can lead to finite Hall conductivity even in absence of an external magnetic field. 

Under the special circumstances of commensuration between the strength of the effective magnetic field (i.e., Berry curvature of magnetic texture) and the itinerant electron density in quasi-2D systems, there is an exciting possibility of obtaining
{\em spontaneous  quantum Hall effect}  (SQHE), i. e.,  quantized Hall response in the absence of magnetic field or uniform spin polarization. 
In a recent work \cite{martin2008}, we have shown that such insulating chiral phase is indeed 
stabilized at zero temperature, $T = 0$, in the weak-coupling regime of a triangular Kondo lattice model (KLM) for a 3/4 filled conduction band ($\rho =3/4$). 
This result was based on the perfect nesting properties of the non-interacting Fermi surface for $\rho =3/4$, which leads to a ``3Q ordering" of classical local moments at $T = 0$, with large uniform scalar spin chirality.

In the present work we explore the stability of chiral order in a 2D Kondo lattice model (KLM) with respect to thermal fluctuations and magnetic field. It is well known that a continuous symmetry cannot be broken at finite temperature in 2D  as long
as the interactions are of short range \cite{mermin1966}. However, a non-coplanar magnetic ordering breaks a discrete $Z_2$ symmetry, in addition to the continuous symmetry, that corresponds to the two disconnected $SO(3)$ sectors of $O(3)$ rotations needed to parametrize the order parameter \cite{villain1977,kawamura1984}. The Ising component of this order parameter (chirality) can survive at finite temperature leading to a chiral spin-liquid: $\langle \chi_{ijk} \rangle \ne 0$ and $\langle {\bf S}_j \rangle = 0$. The interplay between the continuous and the Ising degrees of freedom in such chiral liquid is non-trivial and can lead to deviations from the Ising universality class and a reduction of the transition temperature \cite{momoi1997,domenge2008}.

To study this problem we combined a zero-temperature variational approach with finite-temperature Monte Carlo (MC) simulations of 
the KLM with classical local moments. We demonstrate that the 3Q phase with uniform chirality (UCP) of Ref.  \cite{martin2008} is stable not only at $\rho = 3/4$, but also at $\rho=1/4$ in the intermediate coupling regime $1 \lesssim J/t \lesssim 4$ ($t$ is the hopping amplitude for the conduction electrons and $J$ is the exchange coupling to the localized moments). This result agrees with the small-unit cell variational calculations at $T = 0$ presented in Ref. \cite{akagi2010}. Our MC results provide evidence for a first order thermodynamic phase transition into the UCP. 
The obtained ordering temperature at $J = 2t$, $T_c \simeq 0.03 t$,  is relatively small compared to $J$ and $t$, likely because of the strong frustration and the suppression of the chiral order by the continuous fluctuations of magnetization. Nevertheless, the value of $T_c$ may be high enough for observing SQHE near room temperature 
in transition metal oxides (assuming that $t$ is of order 1eV \cite{dagotto2003}).  Finally, we consider the effect of a magnetic field acting on the local moments both on the stability of the $T = 0$ UCP phase and the spontaneous quantum Hall effect, as well as on the the value of $T_c$.

We consider the Kondo-lattice Hamiltonian on a triangular lattice with periodic boundary conditions,
\begin{eqnarray}
	{\mathcal H}&=&-t\sum_{\langle l,j \rangle \sigma} \left(c^{\dag}_{l\sigma}c_{j\sigma}+{\rm h. c.} \right) -J \sum_{j \mu \nu} {\bf S}_j \cdot c^{\dag}_{j\mu} {\boldsymbol {\sigma}}_{\mu \nu} c_{j\nu},\nonumber
\end{eqnarray}
where $c^{\dag}_{j\sigma}$ $(c_{j\sigma})$ is the creation (annihilation) operator of an electron with spin $\sigma$ on site $j$,
${\bf S}_j$ is a classical Heisenberg spin with $|{\bf S}_j|=1$, 
${\boldsymbol \sigma}_{\mu \nu}=({\sigma}^x_{\mu \nu},{\sigma}^y_{\mu \nu},{\sigma}^z_{\mu \nu})$  is a vector of 
Pauli matrices, and $\langle l,j \rangle$ indicates that $l$ and $j$ are nearest-neighbor sites. Since the  
sign of $J$ is irrelevant for classical moments, ${\bf S}_j$, 
we will assume $J>0$ for concreteness.

\begin{figure}
	\includegraphics[scale =.45, angle=270,trim=0 0 0 0, clip]{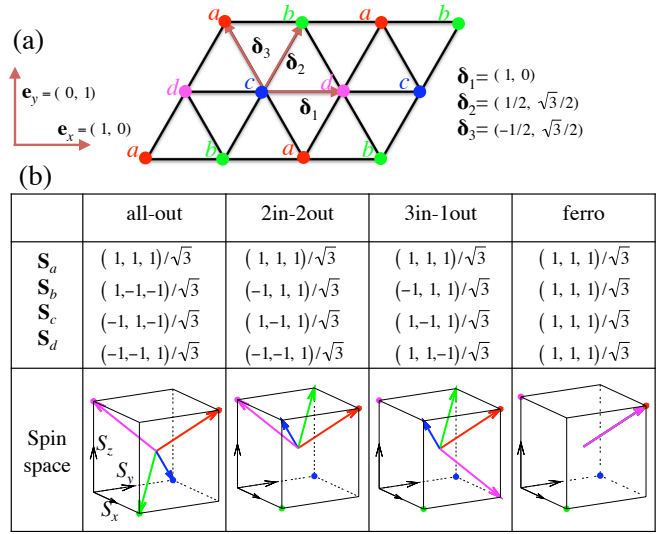}
	\caption{
		(color online) The 4-sublattice spin structures.
		(a) Triangular lattice, with $a-d$ indicating each sublattice
		(b) High-symmetry four-sublattice spin structures. The nomenclature follows the one for magnetic ordering on a tetrahedron with vertices $(abcd)$.}
\label{fig1}
\end{figure}

The state of interest is characterized by the local scalar spin chirality $\chi_{ijk}\equiv {\bm S}_i\cdot{\bm S}_j\times{\bm S}_k$, and its  Fourier transform: 
\begin{equation}
\chi_{\bf q} = N^{-1}   \sum_{\alpha} \chi_\alpha e^{i {\bf q} \cdot {\bf r}}.
\end{equation}
The index $\alpha$ denotes each triangular plaquette, and $N=L^2$ is the total number of lattice sites. The global order parameter for
the UCP is $\langle \chi_{\bf 0} \rangle$. Local spin correlations are described by the 
spin structure factor
\begin{equation}
S({\bf k}) = \frac{1}{N} \sum_{j,l} \langle {\bf S}_l \cdot {\bf S}_j \rangle e^{i{\bf k}\cdot{({\bf r}_l-{\bf r}_j})},
\end{equation}
which is useful for characterization of the $T=0$ spin ordering. We note here that for the perfectly ordered UCP, same as the ``all-out" phase in Fig. \ref{fig1}, simple arithmetic shows that $\chi_{\bf 0}^2 \simeq 0.59$ and $S({\bf 0})=0$, 
while $\chi_{\bf q}^2=0$ and $S({\bf 0})=N$ for the fully polarized ferromagnetic state.

The MC simulation samples the space of all possible classical spin configurations. For each configuration $\{{\bf S}\}$, the electron eigenstates are found by exact diagonalization of $\cal H$.
To study the thermodynamic properties at a fixed filling factor $\rho$, we perform a Legendre transformation of the free energy in the grand canonical ensemble. The resulting free energy is 
\begin{eqnarray}
	F(N,\beta) &\equiv& -\frac{1}{\beta} \ln \Xi,\;\;\;
	\Xi \equiv \sum_{\{\bf S\}}W(\{\bf S\}),\nonumber\\
	W(\{\bf S\})&\equiv&  e^{ -\beta\mu_{\{\bf S\}}N_e  } \prod_{\lambda} \left[ 1+ e^{ -\beta \left\{ \epsilon_\lambda - \mu_{\{\bf S\}} \right\} }  \right]. 
\label{eq:WS}
\end{eqnarray}
The chemical potential 
$\mu_{\{\bf S\}}$ is adjusted for a given configuration $\{\bf S\}$ such that the total number of electrons,
$N_e=\sum_{\lambda} f_\lambda$,  is the same for every spin configuration. Here, 
$\epsilon_\lambda$ is $\lambda$-th eigenvalue of ${\cal H}(\{ \bf S \})$, 
$f_\lambda \equiv {(e^{ \beta\left\{ \epsilon_\lambda - \mu_{\{S\}}\right\} }+1)}^{-1}$, and  $\beta=1/T$. 
In this work we compute the energy density, $\epsilon\equiv N^{-1}\partial_\beta (\beta F)$, and the specific heat, $c\equiv \partial_T\epsilon$.
Our numerical results are based on the single spin flip MC dynamics, with the flip decided by applying the  Metropolis algorithm. We perform 10000 MC sweeps for each processor and estimate the statistical errors using 8 mean values typically.

\begin{figure}
	\includegraphics[scale =.35, angle=270,trim=0 0 0 0, clip ]{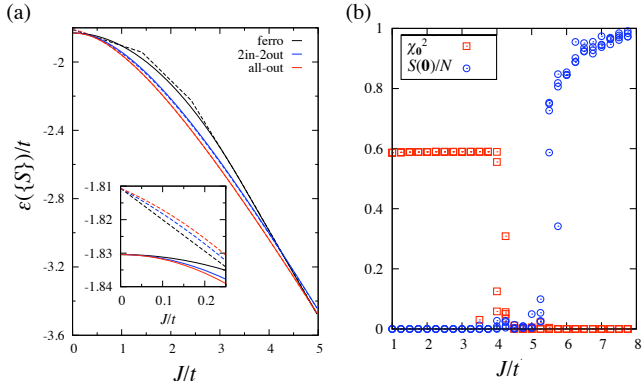}
	\caption{(color online) (a) Energy per site as a function of $J/t$ for
		the all-out, 2in-2out, and ferromagnetic spin structures at $\rho=1/4$ for
		$L=512$ (solid line) and $L=8$ (dashed line).
		Insets show an enlarged view for small $J/t$.
		(b) Coupling $J/t$ dependence of $\chi_{\bf 0}^2$ and $S({\bf 0})$ at quarter filling by Monte-Carlo method with $L=8$ and very low temperature $T/t=10^{-4}$. 
		Four mean values of $\chi_{\bf 0}^2$ and $S({\bf 0})$ are shown for each value of $J/t$ to indicate the lack of convergence the MC simulation in the interval the phase transition, $4 \lesssim J/t \lesssim 6$.
		}
\label{fig2}
\end{figure}

First we consider the $T=0$ case by means of a variational calculation in which we minimize the total energy over several highly symmetric four-sublattice spin structures shown in Fig. \ref{fig1}.
\footnote{See Ref.\cite{akagi2010} for a more exhaustive variational approach.}. 
The ``all-out" and ferromagnetic configurations are the only two  
states that are stabilized as a function of $J/t$ for $\rho=1/4$.
Figure \ref{fig2}(a) includes a comparison between the corresponding  energy  densities $\epsilon (\{{\bf S}\}) \equiv L^{-2}\sum_{\lambda} \epsilon_\lambda f_\lambda$ for $\rho=1/4$. The solid lines are the results for $L=512$. The all-out structure has the lowest energy for $0 < J/t \lesssim 4.9$, while the ferromagnetic configuration becomes the minimum energy state for $J/t \gtrsim 4.9$.
A similar comparison for $L=8$ (dashed lines) indicates that important size effects appear  in the weak coupling regime, $J/t\lesssim1$,
for small values of $N$ (see the inset of Fig. \ref{fig2}(a).)
The energy of the ferromagnetic structure is lower than the energy of the all-out structure for $L=8$ and  small values of $J/t <1$, although this is not true in the thermodynamic limit \footnote{The finite size effects are even more serious for the three-quarter filled case.  The energy of the fully polarized structure is always lower than that the energy of the other two candidates for $L=8$. However, the all-out spin structure becomes the lowest energy state for  $J/t \lesssim 0.6$ and $L> 64$.}.
The all-out structure, which is the lowest energy variational state for $J/t \lesssim 4.9$, has a net uniform scalar spin chirality.  Therefore, it has to produce a spontaneous Hall effect. By explicit calculation of the Hall conductivity from the electron band structure, we find the quantized value $\sigma_{xy}= \pm e^2/h$ for $0.7 \lesssim  J/t  \lesssim 4.9$.

The MC calculations at low temperatures allow us to test the results of the variational approach, not being limited to the $2\times2$ magnetic unit cell size.
Figure \ref{fig2}(b) shows our MC results for the $J/t$ dependence of $\chi_{\bf 0}^2$ and $S({\bf 0})$ at 
a very low temperature $T/t=10^{-4}$ and $L=8$. These results show that the all-out spin structure with uniform scalar 
chirality is indeed stable for $J/t \lesssim 4$, while the ferromagnetic phase is stabilized for $J/t \gtrsim 5$. 
Our MC simulation does not converge at these very low temperatures in the interval $4 \lesssim J/t \lesssim 6$. 
We traced the lack of convergence to an intervening spiral phase, which in the $L = 8$ systems is slightly lower in energy than the fully polarized and the all-out phases for $4.9\lesssim J/t\lesssim6.3$. However, for larger systems ($L >64$), the spiral state  becomes unstable.
Therefore, we conclude that the non-convergent window between the all-out and the fully polarized  phases is only a finite size effect, and there is a direct transition from the all-out to the ferromagnetic phase as a function of $J$ \cite{akagi2010}. We also find that $S({\bf 0})$ is finite while $\chi_{\bf 0}^2$ is almost zero below $J/t \sim 0.5$. This is a clear consequence 
of the finite size effect that was already discussed in our variational calculation [see Fig.\ref{fig2}(a)]. The intermediate coupling range, $1 <J/t<4$, appears to be stable against size effects.

\begin{figure}
	\includegraphics[scale =.5, angle=270,trim=0 0 0 0, clip ]{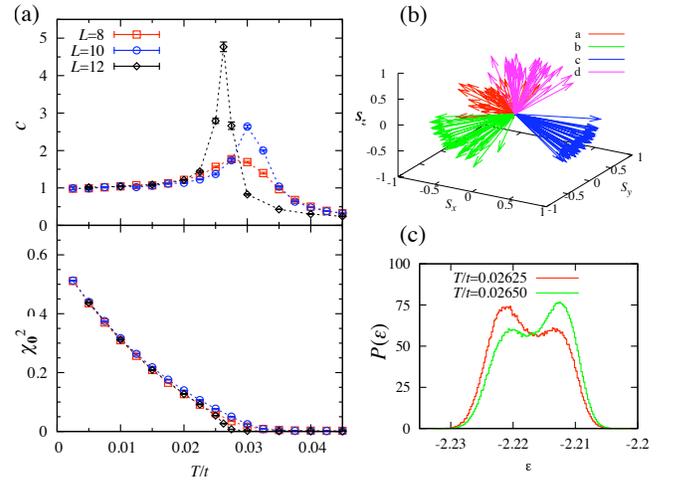}
	\caption{
		(color online) Results of Monte-Carlo simulations at quarter filling and $J/t=2$.
		(a) Temperature dependence of specific heat $c$  and $\chi_{\bf 0}^2$.
		(b) Snapshot of a spin configuration. Each color corresponds to each of the four different sublattices ($L=12$, $T/t=0.005$).
		(c) Internal energy distribution near the critical temperature for $L=12$.
	}
\label{fig3}
\end{figure}

We now turn to the question of the finite temperature stability of the chiral magnetic phase. Our main results are presented in Fig \ref{fig3}.  To control finite size effects, we studied several system sizes $L$ at fixed coupling strength $J/t=2$. This value of $J/t$ gives a robust all-out chiral ordering at $T = 0$. As expected from the simple Ising argument in 2D, this chiral ordering should persist for a finite range of temperatures.
Starting from the maximum possible value, the scalar chirality decreases as a function of temperature, vanishing at $T_c \simeq 0.026 t$.  The snapshot of a spin configuration at $T  = 0.005t$ in Fig. \ref{fig3}(b) illustrates how the finite-temperature fluctuations destabilize the chiral ordering. The calculated specific heat curve exhibits a
very sharp and symmetric peak at $T_c$, which is an indication of a first order phase transition.  To test this possibility, we analyzed the temperature dependence of the internal energy probability distribution function, Fig. \ref{fig3}(c). We found that it has a bimodal distribution for $T$ near $T_c$ and $L=12$. In combination with the specific heat behavior, this result indicates that the corresponding thermodynamic phase transition between the paramagnetic and the chiral all-out state is of the first order.

Magnetic field, ${\bf H}$, is an important parameter that can be used to control the chiral magnetic states. It couples to both local and itinerant electron spins, as well as to the orbital motion of electrons, and hence can lead to rich variety of phases. For simplicity, here we consider the minimal -- Zeeman --coupling to the local moments
\begin{equation}
{\cal H}_z = - g \mu_B \sum_{j} {\bf H} \cdot {\bf S}_j,
\end{equation} 
with ${\bf H}=H{\hat {\bf z}}$, 
and neglect the Zeeman and orbital coupling to the conduction electrons. This approximation is rigorously justified when the Hund's coupling is much smaller than the itinerant electron bandwidth. In this case, the electron Pauli susceptibility can be neglected relative to the local moment susceptibility, while the orbital effect of Berry curvature induced by the magnetic texture dominates the orbital effect of the applied  field.

\begin{figure}[!htb]
	\includegraphics[scale =.45, angle=270,trim=0 0 0 0, clip ]{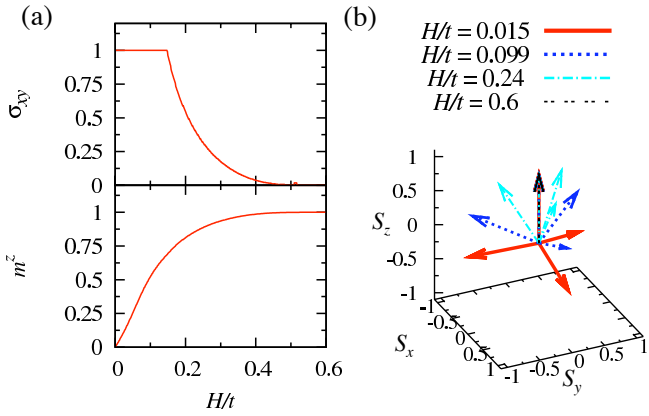}
	\caption{(color online) Variational calculation of the ground state of ${\cal H}+{\cal H}_z$ by minimizing the energy
	over all possible four-sublattice spin configurations ($L=2048$): a) Hall conductance, $\sigma_{xy}$, and uniform magnetization along
	the field direction, $m^z$, as a function of the applied magnetic field $H/t$. (b) Field dependence of the 
	four-sublattice spin structure of the variational ground state.}
	\label{fig4}
\end{figure}

The  $T=0$ variational approach for $J/t=2$ reveals that the  all-out structure  is distorted by the field in the way that interpolates between the all-out and three-in one-out structure [Fig. \ref{fig1}(b)], until it saturates at $H_{sat} \simeq 0.5t$ as shown 
in Fig. \ref{fig4}(b). The spins of one of the four sublattices are always aligned with the field direction.
The gap in the electron spectrum closes at the critical field $H_c/t \simeq 0.15$ leading to an insulator-to-metal transition. Correspondingly, $\sigma_{xy}$, starts to decrease continuously at $H = H_c$, from its quantized value for $H\leq H_c$
to zero for $H\geq H_{sat}$. In this way, the application of an external field suppresses SQHE  in the insulating state by inducing a metallic phase with anomalous Hall effect. 

\begin{figure}[!htb]
	\includegraphics[scale =.5, angle=270,trim=0 0 0 0, clip ]{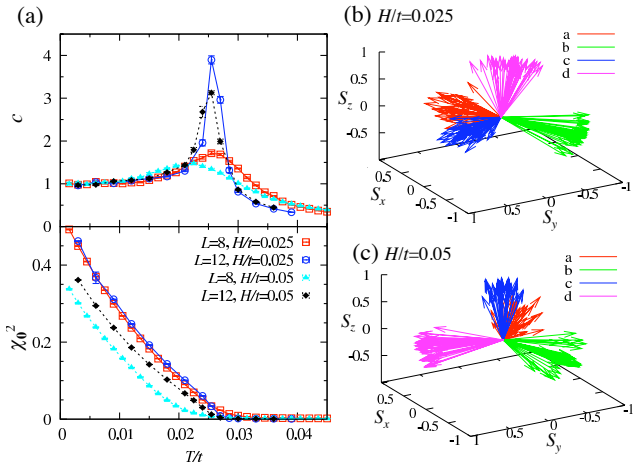}
	\caption{
		(color online) Results of Monte-Carlo simulation at quarter filling and $J/t=2$, for two values of applied magnetic field, $H/t=0.025$ and 0.05,
		(a) temperature dependence of specific heat $c$ and $\chi_{\bf 0}^2$ ,
		(b,c) configurations of all spins classified according to sublattice ($L=12$, $T/t=0.003$), with
		(b) $H/t=0.025$,
		(c) $H/t=0.05$.
	}
\label{fig5}
\end{figure}

We performed MC simulations in the low field region, $H=0.025t$ and $H=0.05t$, for $L=8$ and $L=12$ (for larger values of $H$ the variational approach revealed spurious finite size effects in systems with $L < 64$).  
The results are shown in Fig. \ref{fig5}. The position of the specific heat peak (Fig. \ref{fig5}(a)) and the 
onset of the chiral order parameter (Fig. \ref{fig5}(a)) indicate that the transition temperature to the chiral phase 
is suppressed by the presence of the magnetic field. The snapshots of the spin configurations in each sublattice (identified with color) show a result that is consistent with the variational approach: the spins are canted towards the three-in one-out configuration, with one of the spin sublattices tending to be aligned with the applied field (see Figs. \ref{fig5}(b) and (c)). The fluctuations around this preferred orientation decrease with increasing $H$.

In summary, we have shown that a chiral spin liquid is present in the quarter-filled  KLM with classical local moments on a triangular lattice.  This liquid exhibits spontaneous quantum Hall effect, which can be tuned by an external magnetic field. The phase is stable with respect to thermal fluctuations and exists in a wide range of the Hund's coupling strengths.  It can therefore be realized, e.g., in manganese-based materials \cite{dagotto2003}.  
It may also be relevant to Na$_x$CoO$_2$ \cite{martin2008, li2010} in the limit of quantum local moment, $S = 1/2$.

We thank S. Nakatsuji and Y. Motome for useful discussions.
This work was carried out under the auspices of the NNSA
of the U.S. DOE  at LANL under Contract No. DE-AC52-06NA25396 and supported by the LANL/LDRD Program.

\bibliography{apssamp}

\end{document}